# 4.5 GHz Lithium Niobate MEMS Filters with 10% Fractional Bandwidth for 5G Front-ends

Yansong Yang, *Student Member, IEEE*, Ruochen Lu, *Student Member, IEEE*, Liuqing Gao, *Student Member, IEEE*, and Songbin Gong, *Senior Member, IEEE*

*Abstract*— This paper presents a new class of micro-electro-mechanical system (MEMS) C-band filters for 5G front-ends. The filter is comprised of resonators based on the first-order asymmetric Lamb wave (A1) mode in thin film lithium niobate. Two filters have been demonstrated at 4.5 GHz with sharp roll-off, flat in-band group delay, and spurious-free response over a wide frequency range. The first design shows a fractional bandwidth (FBW) of 10%, an insertion loss (IL) of 1.7 dB, an out-of-band (OoB) rejection of -13 dB, and a compact footprint of 0.36 mm$^2$, while the second design shows an FBW of 8.5%, an IL of 2.7 dB, an OoB rejection of -25 dB, and a footprint of 0.9 mm$^2$. The demonstrations herein mark the largest fractional bandwidth (FBW) achieved for acoustic-only filters at 5G frequencies.

*Index Terms*—5G front-ends, New radio, C-band, MEMS, Acoustic, Resonators, Filters, lithium niobate, Spurious modes suppression, Constant in-band group delays.

## I. INTRODUCTION

IN response to the explosion of mobile data from video streaming, virtual reality, and Internet-of-Things (IoT), the wireless industry has moved towards 5G to overcome the limitations of existing wireless networks. To balance the needs for wide-area coverage and high data rates, 5G new radio (NR) has been proposed in C-band (4-8 GHz) with a significant increase in fractional bandwidths (FBWs) that can be as high as 13% [1], [2]. Such a large FBW challenges the capabilities of the incumbent RF front-end acoustic filters of which the FBWs remain less than <5%. The constraints on FBWs fundamentally arise from the electromechanical coupling ($k_t^2$) that can be achieved for an adopted mode in a given material. For instance, state-of-the-art acoustic filters based on lithium tantalite (LiTaO$_3$) or lithium niobate (LiNbO$_3$) surface acoustic wave (SAW) resonators and aluminum nitride (AlN) bulk acoustic wave (FBAW) resonators only have $k_t^2$<10%. Thus, it is challenging to support the emerging 5G NR with more than 5% FBW [3]–[5] without resorting to lumped-element-based bandwidth-widening techniques [6].

To overcome the bottleneck of limited $k_t^2$, one promising solution is the recently emerged first-order asymmetric lamb wave (A1) based on Z- and Y-cut LiNbO$_3$ [7]–[10]. The A1 resonators in Z-cut LiNbO$_3$ have been demonstrated with a $Q$ of 527 and $k_t^2$ of 29% at 5 GHz, which are sufficient to meet the BW requirements for 5G bands in the Sub-6 GHz range. However, the spurious modes in these past demonstrations remain a major bottleneck [11], and high-performance A1 filters have not been demonstrated.

In this study, we aim to demonstrate C-band LiNbO$_3$ filters based on the A1 mode with large FBW for the first time. To this end, the spurious modes suppression and resonator frequency offset are analyzed. To validate our designs, the devices were fabricated on a 500 nm thick Z-cut LiNbO$_3$ thin film. The fabricated resonator has shown a $k_t^2$ of 28% and a $Q$ of 430. The fabricated filters have been demonstrated with large FBWs (>8.5%), small insertion loss (IL<2.7 dB), flat in-band group delays and spurious-free responses.

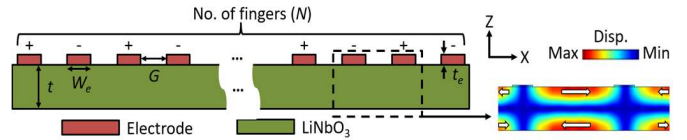

Fig. 1. Configuration of a Z-cut LiNbO$_3$ A1 mode resonator with displacement mode shape (Arrows denote the displacement directions).

## II. DEVICES DESIGN

As seen in Fig. 1, the A1 mode can be efficiently excited in a Z-cut LiNbO$_3$ thin film with top-only interdigital transducers (IDTs). However, the electric field excited by top-only IDTs within the thin film can have components that couple to the unwanted spurious modes near the A1 mode [12]. To mitigate the spurious modes, the E-field distribution has been optimized based on the analysis described in our previous work by using an optimal electrode gap [12].

With the optimized resonators, a simple ladder topology consisting of series and shunt resonators is adopted for the filter demonstrations. As shown in Fig. 2, two designs using identical series and shunt resonators but with the different filter orders are constructed with tradeoff between the insertion loss and out-of-band rejection. To make the filter footprint compact and symmetric, each shunt resonator is implemented with two identical resonators in parallel. The resonant frequencies of series and shunt resonators are designed with a frequency offset between them, which approximately gives the bandwidth of the filter [13]. To achieve the large bandwidth allowed by the $k_t^2$ of the standalone resonators, the offset should approach the spectral separation between the series and anti-resonances. The required resonant frequency offset can be attained by varying the electrode separation ($G$) as suggested by the dispersion of A1 shown in Fig. 3(a) [8]. This technique permits the monolithic implementation of multi-frequency resonators required by the ladder topology as well as lithography-based frequency trimming.

Based on the above analyses, the series and shunt resonators for C-band acoustic filters with 10% FBW are designed with

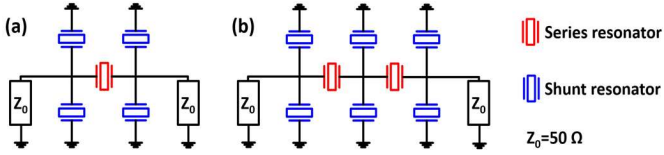

Fig. 2. Ladder filters of (a) Design A and (b) Design B.

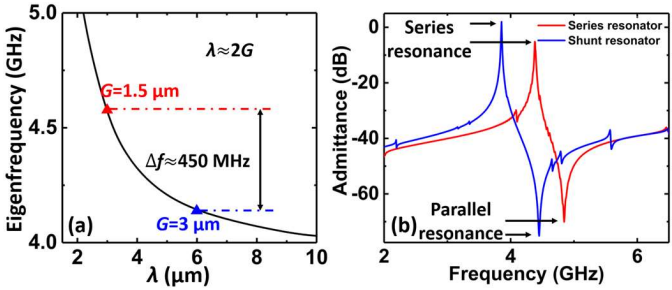

Fig. 3. (a) Dispersion curve of the A1 mode in a 500 nm thick Z-cut LiNbO$_3$ thin film. (b) COMSOL simulated admittance responses of the series and shunt resonators.

TABLE I
Design Parameters of the Series and Shunt Resonators

| Parameter | $t$ (nm) | $W_e$ (μm) | $G$ (μm) | $L$ (μm) | $t_e$ (nm) | $N$ | $C_0$ (fF) |
|---|---|---|---|---|---|---|---|
| Series resonator | 500 | 0.5 | 1.5 | 80 | 50 | 50 | 380 |
| Shunt resonator | 500 | 1.5 | 3 | 80 | 50 | 80 | 320 |

*parameters are marked and defined in Fig. 1.

the key parameters listed in Table I. A thickness ($t$) of 500 nm is chosen to predominately set the filter center frequency to 4.5 GHz. The electrode gaps ($G$) are set to 1.5 and 3 μm for series and shunt resonators to suppress most significant spurious modes [10] while providing the required frequency offset of 450 MHz for achieving the 10% FBW [as shown in Fig. 3(a)]. In addition to selecting optimal values for $G$, an optimal ratio of $W_e$ to $G$ is also required to mitigate spurious modes without inducing large electrical loss from an excessively small $W_e$. To balance the loss and spurious considerations, $W_e$ is set to be 0.5 μm and 1.5 μm respectively for series and shunt resonators. Last, to attain an adequate $C_0$ for matching to 50 Ω and sufficient out-of-band rejections, the numbers of electrodes are increased to 50 and 80 respectively for series and shunt resonators, and the lengths of the resonators are set to 80 μm. As seen in Fig. 3 (b), the COMSOL simulated responses of the designed series and shunt resonators confirm an intended frequency offset of 450 MHz with damped spurious responses.

## III. EXPERIMENTAL RESULTS AND DISCUSSIONS

To demonstrate the designs, the optimized resonators and filters were fabricated on a 500 nm transferred Z-cut LiNbO$_3$ thin film following the process described in [9]. The effectiveness of our optimized design on spurious mitigation is highlighted via a comparison between measured responses of an un-optimized and an optimized A1 mode resonator [Fig. 4(a) and (b)]. The optimized design shows notable suppression of spurious modes with both $k_t^2$ and $Q$ significantly enhanced. The measured response of a fabricated filter without applying spurious suppression is included in Fig. 4(c) to indicate that unattended spurious modes lead to the deep in-band ripples and out-of-band responses.

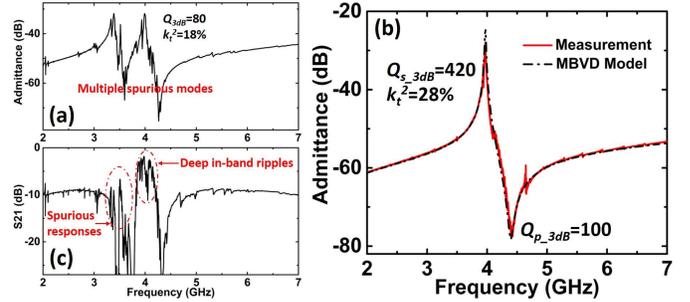

Fig. 4. Measured performance of an (a) un-optimized A1 resonator with multiple significant spurious modes and the (b) optimized A1 resonator with spurious modes damped and simultaneously enhanced $k_t^2$ and $Q$. (c) Measured performance of a filter comprised by spurious-laden resonators.

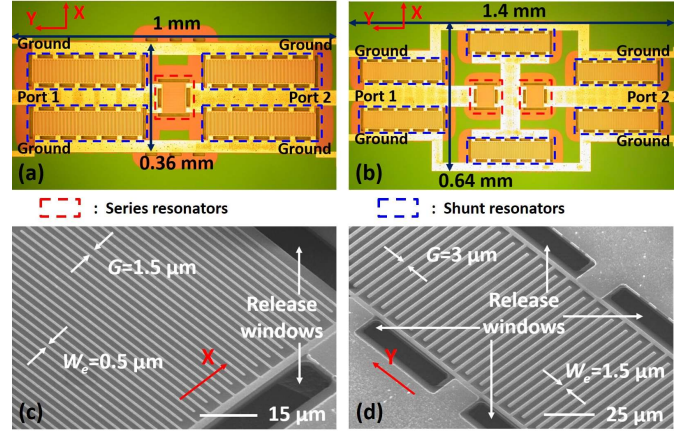

Fig. 5. Optical microscope images of (a) Design A and (b) Design B. SEM images of the (c) series resonator and (d) shunt resonator.

The optical microscope images of the two fabricated filters, Design A and Design B, are shown in Fig. 5(a) and (b). As the A1 mode has negligible difference in propagation characteristics along X and Y axes, the series and shunt resonators are oriented along X- and Y-axis respectively to attain a compact footprint without degrading performance. The overall footprint for these two acoustic designs are 0.36×1 mm$^2$ and 0.64×1.4 mm$^2$ respectively. The electrode separation and width of the series and shunt resonators are labeled in the zoom-in SEM images [Fig. 5(c) and (d)].

The measured S21 and S11 of these two fabricated filters are shown in Fig. 6(a). Design A shows an IL of 1.7 dB, an FBW of 10%, and an out-of-band rejection of -13 dB, while the Design B shows an IL of 2.7 dB, an FBW of 8.5%, and an out-of-band rejection of -25 dB. For Design A and B, the 4 dB-FBWs, defined as the fractional frequency range over which IL is below 4 dB, are 8.7% and 6% respectively. As shown in Fig. 6(b), the magnitude of in-band IL ripples is only around 0.5 dB, which is much smaller than the measured results shown in Fig. 4(b) due to the spurious mode suppression. Additionally, a flat in-band group delay is also achieved. As shown in Fig. 6(c), the group delay variations in the entire passband are around 3 ns which is sufficiently small to avoid RF signal distortion.

To analyze the performance and identify the venues for further improvements, an equivalent circuit model shown in Fig. 7(a) is constructed to interpret the measurement result. Consistent with the conclusion in [14], the electromagnetic (EM) effects are accounted for via including the surface resistance ($R_s$) and inductance ($L_{s\_series}$ and $L_{s\_shunt}$) associated with electrodes and leading lines. $C_p$ is added as the feedthrough capacitance to account for the parasitics. Performance of each fabricated series or shunt resonator in the filter is also taken into consideration for constructing the equivalent circuit model. According to the statistical results of the $Q$s of the fabricated standalone resonators, which range from 150 to 420, it can be assumed that the $Q$s of the series and shunt resonators in the measured filters are on an average around 200. As shown in Fig. 7(b), the simulated response based on the equivalent circuit model exhibits excellent agreement with the measurement, and the key parameters are listed in Table III. Based on the above analysis, we believe that a further optimized fabrication process that can increase the yield of high-performance resonators is the key to higher performance filters. In addition, the designs with smaller $R_s$, $L_{s\_series}$, $L_{s\_shunt}$, and $C_p$ can also be proven to be helpful.

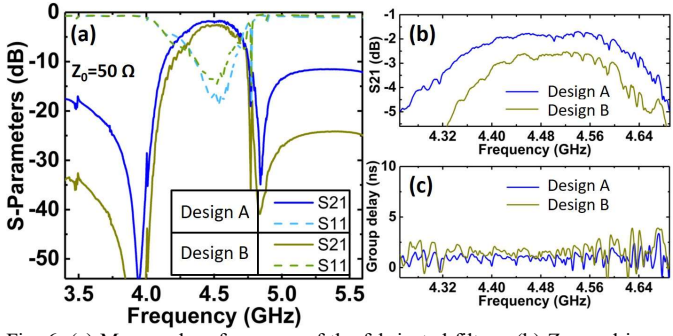

Fig. 6. (a) Measured performance of the fabricated filters. (b) Zoomed-in response of the intended passbands. (c) Measured in-band group delay.

TABLE II
Key Measured Parameters of the Fabricated Filters

| Parameter | $f_0$ | FBW | IL | OoB rejection | In-band ripples | Group delay |
|---|---|---|---|---|---|---|
| Design A | 4.5 GHz | 10% | 1.7 | <-13 dB | <0.5 dB | <3 ns |
| Design B | 4.5 GHz | 8.5% | 2.7 | <-25 dB | <0.7 dB | <4 ns |

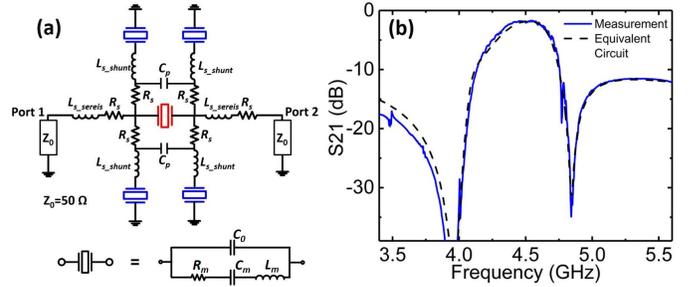

Fig. 7. (a) Equivalent circuit model accounting for EM effects. (b) Comparison between the measured and simulated responses based on the equivalent circuit mode.

TABLE III
Key Parameters of the Equivalent Circuit

| Parameter | $Q$ | $R_s$ | $L_{s\text{-series}}$ | $L_{s\text{-shunt}}$ | $C_p$ | IL |
|---|---|---|---|---|---|---|
| Value | 200 | 4 Ω | 0.2 nH | 0.3 nH | 15 fF | 1.7 dB |

## IV. CONCLUSION

In this work, a new class of C-band acoustic filters based on the first-order asymmetric lamb wave mode in Z-cut LiNbO$_3$ is presented. This work marks the largest demonstrated fractional bandwidth for acoustic-only filters at 5G frequencies, showing the strong potential of lithium niobate A1 filters for 5G RF front-ends.